\documentclass[10pt,letterpaper,twocolumn]{article}
\usepackage{ol2}

\usepackage{graphicx}
\begin{document}
\twocolumn[ 
\title{Two-color surface lattice solitons}

\author{Zhiyong Xu$^{*}$ and Yuri S. Kivshar}

\address{Nonlinear Physics Center and Center for Ultra-high bandwidth
Devices for Optical Systems (CUDOS), \\ Research School of Physical
Sciences and Engineering, The Australian National University, Canberra ACT 0200, Australia\\
$^*$Corresponding author: xzy124@rsphysse.anu.edu.au}

\begin{abstract}
We study the properties of surface solitons generated at the edge of a semi-infinite photonic
lattice in nonlinear quadratic media, namely two-color surface lattice solitons. We analyze
the impact of phase mismatch on existence and stability of surface modes, and find
novel classes of two-color twisted surface solitons which are stable in a large domain of their
existence.
\end{abstract}

] 

\maketitle

Surface modes appear as a special type of waves
localized near an interface separating two different
media. In optics, linear electromagnetic surface waves are known to exist
at an interface separating homogeneous and periodic dielectric media~\cite{Yeh_APL_78}.
Recently, the interest in the study of electromagnetic surface waves has
been renewed, and it was shown theoretically~\cite{OL_george,OL_molina,PRL_kartashov} and
experimentally~\cite{PRL_george,OE_stegeman,smirnov,PRL_canberra} that
nonlinearity-induced self-trapping of light may become possible near the edge of a
one-dimensional waveguide array leading to the formation of discrete surface solitons
(see also the review papers~\cite{review_1,review_2}). In particular, it was found that
the self-trapped surface modes acquire some novel properties
different from those of the discrete solitons in infinite lattices: discrete surface states
can only exist above a certain threshold power and, for the same value of the power, up to
two different surface modes can exist simultaneously. This can be understood as discrete optical
solitons~\cite{book} localized near the surface and the action of a
repulsive force from the surface~\cite{OL_molina}.

Surface solitons are usually studied for cubic or saturable nonlinear media. However, Siviloglou
et al.~\cite{OE_stegeman} reported on the first observation of
discrete quadratic surface solitons in periodically poled lithium
niobate waveguide arrays. By operating on either side of the phase-matching
condition and using the cascading nonlinearity, they observed both in-phase and staggered
discrete surface solitons. They also employed a discrete model with decoupled waveguides
at the second harmonics to model some of the effects observed experimentally.

The purpose of this Letter is two-fold. First, we extend substantially the earlier theoretical
analysis performed by Siviloglou et al.~\cite{OE_stegeman} and study, for the first
time to our knowledge, two-color quadratic surface solitons in  a continuum model with a truncated
periodic potential. We analyze the effect of the mismatch on the existence, stability, and generation
of surface solitons located at the edge of a semi-infinite waveguide array in nonlinear quadratic media. Second,
we reveal the existence of novel classes of surface solitons which are stable in a large domain of their
existence.

\begin{figure}[t]
\begin{center}
\includegraphics[width=7cm,bb=185 365 437 599]{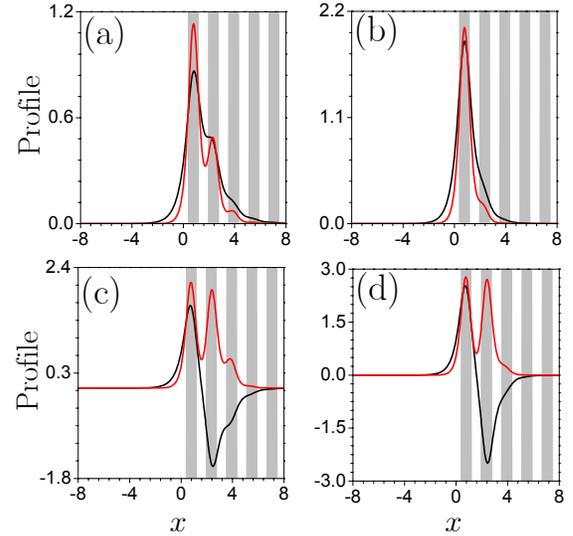}
\end{center}
\caption{(Color online) Profiles of two-color odd surface
solitons with (a) $b_{1}=1.5$ and (b) $b_{1}=2$. Profiles of
two-color twisted surface solitons with (c) $b_{1}=1.75$ and (d)
$b_{1}=2.2$. Black and red curves show the profiles of FF and
SH fields, respectively. Lattice depth $p=1$, and phase mismatch
$\beta=0$. In the white regions $R(x)<1$, while in the gray
regions $R(x)\geq1$.} \label{figure1}
\end{figure}

We consider propagation of light at the interface of semi-infinite
lattice imprinted in quadratic nonlinear media, which involves
the interaction between fundamental frequency (FF) and
second-harmonic (SH) waves. Light propagation is
described by the following coupled nonlinear equations \cite{Kartashov1,Xu}
\begin{eqnarray}\label{eq:model}
&& i\frac{\partial q_{1}}{\partial
z}=\frac{d_{1}}{2}\frac{\partial^{2}q_{1}}{\partial x^{2}}-
q_{1}^{*}q_{2}\texttt{exp}(-i\beta z)-pR(x)q_{1} \nonumber \\
&&i\frac{\partial q_{2}}{\partial
z}=\frac{d_{2}}{2}\frac{\partial^{2}q_{2}}{\partial x^{2}}-
q_{1}^{2}\texttt{exp}(i\beta z)-2pR(x)q_{2}.
\end{eqnarray}
where $q_{1}$ and $q_{2}$ represent the normalized complex amplitudes
of the FF and SH fields, $x$ and $z$ stand for the normalized
transverse and longitudinal coordinates, respectively, $\beta$ is
the phase mismatch, and $d_{1}=-1$, $d_{2}=-0.5$; $p$ is the
lattice depth; the function $R(x)=0$ at $x<0$ and
$R(x)=1-\texttt{cos}(K x)$ at $x\geq 0$ describes the profile
of a truncated periodic lattice with modulation $K$. The system of
Eq.~(\ref{eq:model}) admits several conserved quantities including
the power $P=\int_{-\infty}^{\infty}(|q_{1}|^{2}+|q_{2}|^{2})dx$.

The stationary solutions for the lattice-supported surface solitons
can be found in the form
$q_{1,2}(x,z)=u_{1,2}(x)\texttt{exp}(ib_{1,2}z)$, where $u_{1,2}(x)$
are real functions, and $b_{1,2}$ are real propagation constants
satisfying $b_{2}=\beta+2 b_{1}$. Families of surface solitons are determined by
the propagation constant $b_{1}$, the lattice depth $p$, and the phase mismatch $\beta$.
For simplicity, we set the modulation parameter $K=4$. To analyze stability we examine perturbed solutions
$q_{1,2}(x,z)=[u_{1,2}(x)+U_{1,2}(x,z)+iV_{1,2}(x,z)]\texttt{exp}(ib_{1}z)$, where real parts $U_{1,2}$
and imaginary parts $V_{1,2}$ of perturbation can grow with complex rate $\delta$. Linearization
of Eq.~(\ref{eq:model}) around $u_{1,2}$ yields the eigenvalue problem
\begin{eqnarray}\label{eq:model2}
&& \delta U_{1}=\frac{d_{1}}{2}\frac{\partial^{2}V_{1}}{\partial x^{2}}-
(u_{1}V_{2}-u_{2}V_{1})-pRV_{1}+b_{1}V_{1} \nonumber \\
&& \delta V_{1}=-\frac{d_{1}}{2}\frac{\partial^{2}U_{1}}{\partial x^{2}}+
(u_{1}U_{2}+u_{2}U_{1})+pRU_{1}-b_{1}U_{1} \nonumber \\
&& \delta U_{2}=\frac{d_{2}}{2}\frac{\partial^{2}V_{2}}{\partial x^{2}}-
2u_{1} V_{1}-2pRV_{2}+b_{2}V_{2} \nonumber \\
&& \delta V_{2}=-\frac{d_{2}}{2}\frac{\partial^{2}U_{2}}{\partial x^{2}}+
2u_{1}U_{1}+2pRU_{2}-b_{2}U_{2}.
\end{eqnarray}
which we solve numerically and find the growth rate $\delta$.

\begin{figure}[t]
\begin{center}
\includegraphics[width=7cm, bb=192 387 434 610]{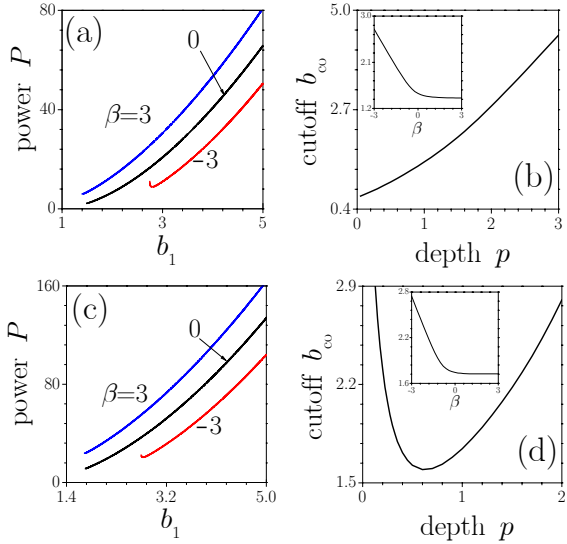}
\end{center}
\caption{(Color online)(a) Power of surface solitons vs. propagation constant
for different phase mismatches. (b) Cutoff of the propagation constant $b_1$ for surface solitons as a
function of depth at $\beta=0$, Inset shows the cutoff vs. phase mismatch at $p=1$.
(c) Power of twisted surface solitons vs. $b_1$ for different phase mismatches.
(d) Cutoff of the propagation constant for twisted surface solitons as a function of lattice
depth at $\beta=0$. Inset shows the cutoff
vs. the phase mismatch at $p=1$.}
\label{figure2}
\end{figure}

\begin{figure}[t]
\begin{center}
\includegraphics[width=7cm, bb=161 486 455 596]{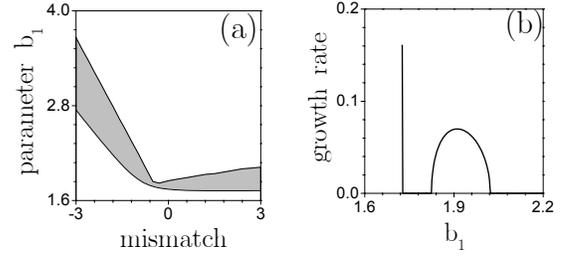}
\end{center}
\caption{(a) Instability domain (shaded) for
twisted surface solitons on the $(b_{1},\beta)$ plane at $p=1$.
(b) Real part of the instability growth rate for twisted surface
solitons vs. propagation constant at $\beta=0$ and
$p=0.5$.} \label{figure3}
\end{figure}

Figures~\ref{figure1}(a,b) show the examples of the simplest two-color
surface lattice solitons, the so-called odd solitons. Such surface modes reside in the first
lattice channel, where both FF and SH fields reach their maxima. Due to
the lattice truncation, the solitons have asymmetric profiles at
lower power [Fig.~\ref{figure1}(a)]. There exists a lower cutoff ($b_{\texttt{co}}$) of the propagation
constant for the existence of odd solitons. The power of odd solitons is a nonmonotonic function of the
propagation constant, and there is a narrow region
close to the cutoff where the power dependence changes its slope, $dP/db_{1}<0$ [Fig.~\ref{figure2}(a)].
We find that the cutoff $b_{\texttt{co}}$ is a monotonically increasing function of lattice depth $p$,
and it is a decreasing function of the phase mismatch [Fig.~\ref{figure2}(b)]. It should be noted that
the critical power for the existence of two-color surface lattice solitons depends on the phase mismatch,
it reaches the minimum value for the exact phase matching ($\beta=0$). Linear stability analysis reveals
that these odd surface solitons are stable almost in the whole domain of their existence except a very narrow
region near the cutoff $b_{co}$ where the exponential instability develops. Direct numerical simulations of
the model~(\ref{eq:model}) confirm the results of the linear stability analysis.

In addition to the simplest solitons described above, we find
various families of higher-order surface lattice solitons, which can be viewed as
the combination of several in-phase and out-of-phase odd solitons. Here we only focus on
the case where the FF field features the out-of phase combinations because the modes
with in-phase combinations are unstable. Being reminiscent to its discrete counterparts
such as twisted localized modes, here we term the mode with out-of phase combination as
\emph{twisted surface lattice soliton}. Typical profiles of twisted modes residing at
the edge of a semi-infinite lattice are shown in Figs.~\ref{figure1}(c,d).
Similar to the properties of odd solitons, the power of twisted surface solitons
is a nonmonotonic function of the propagation constant, and there exists a narrow region close
to the cutoff where $dP/db_{1}<0$ [Fig.~\ref{figure2}(c)]. However, one should note that
the cutoff $b_{\texttt{co}}$ is a nonmonotonic function of the lattice depth $p$, as shown in Fig.~\ref{figure2}(d).
Likewise, the critical power for the existence of twisted solitons reaches the minimum value at $\beta=0$ .

Importantly, we find that the twisted surface lattice solitons become
completely stable when the power exceeds a certain threshold value [Fig.~\ref{figure3}(a)],
while the instability domain expands with the growth of the absolute value of phase
mismatch $\beta$. Figure~\ref{figure3}(b) shows the real part of the perturbation growth
rate versus the propagation constant, where the left part corresponds to the
exponential instability, while in the right part solitons suffer from oscillatory instabilities.
Results from linear stability analysis are confirmed by the direct numerical simulation of the model~(\ref{eq:model}).

\begin{figure}[t]
\begin{center}
\includegraphics[width=7cm, bb=190 369 437 616]{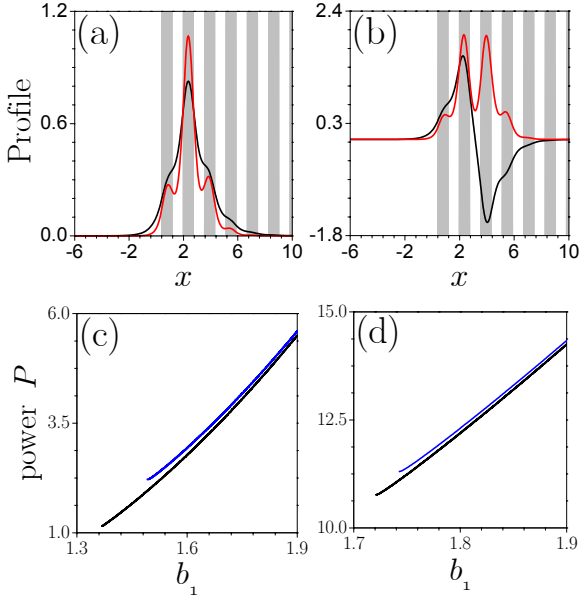}
\end{center}
\caption{(Color online) Profiles of two-color surface solitons in the second channel:
(a) odd one with $b_{1}=1.5$ and (b) twisted one with $b_{1}=1.75$. The black and red curves show
the profiles of FF and SH fields, respectively. Power of (c) odd and (d) twisted surface solitons in the first (blue)
and second (black) channel of the lattices. Here lattice depth $p=1$, and phase matching $\beta=0$.} \label{figure6}
\end{figure}

We also study the properties of two-color surface modes which are shifted from the lattice edge,
similar to the analysis performed earlier for the discrete cubic model~\cite{OL_molina}. Figures~\ref{figure6}(a,b)
show some illustrative examples of odd and twisted states residing in the second channel of the semi-infinite
waveguides from the interface. Such modes require much lower critical power comparing with surface modes residing
at the interface [Fig.~\ref{figure6}(c,d)]. Linear stability analysis reveals that two-color surface solitons in the second
channel are stable when their power exceeds a certain threshold.

\begin{figure}[t]
\begin{center}
\includegraphics[width=7cm, bb=227 528 380 685]{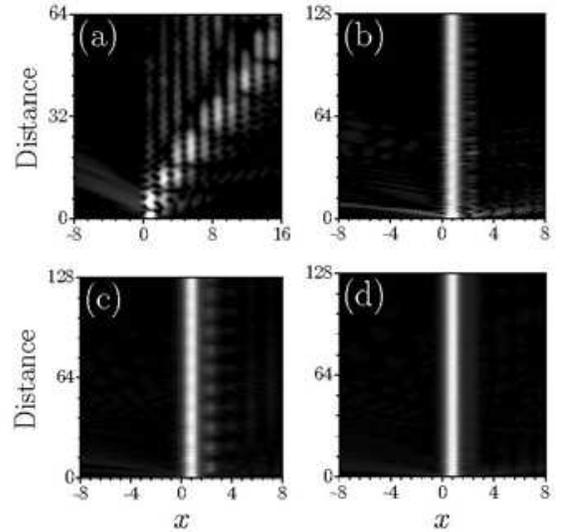}
\end{center}
\caption{Excitation of two-color surface solitons from Gaussian input beams:
(a) $a_{1}=1.5$, $a_{2}=0$; (b) $a_{1}=2.8$, $a_{2}=0$; (c) $a_{1}=1.5$,
$a_{2}=1$; and (d) $a_{1}=2$, $a_{2}=2.5$. Here (a-c) $\beta=0$, and (d) $\beta=-3$.
In all cases, $p=1$ and only the SH field is shown.} \label{figure5}
\end{figure}

To address the issue of excitation of two-color surface solitons, we perform a comprehensive study
of the soliton generation by two Gaussian beams $q_{1,2}=a_{1,2}\texttt{exp}(-x^{2})$.  As shown in Figs.~\ref{figure5}(a-d),
for lower powers the beams just experience repulsion from the surface and diffraction [(a)]. For high enough powers,
surface solitons can be excited either from only FF field [(b)] or both  FF and SH fields [(c)] with a proper phase matching. Two-color
surface solitons can be formed for other conditions ($\beta=-3$), but for higher input powers [(d)].

In conclusion, we have analyzed the existence, stability, and generation of two-color quadratic surface
solitons in  a continuum model with a truncated periodic potential, and also revealed the existence
of novel classes of stable parametrically coupled surface states.

We acknowledge a support of the Australian Research Council.

\end{document}